\documentclass[]{spie}

\usepackage{amsmath,amsfonts,amssymb}
\usepackage{graphicx}
\usepackage[colorlinks=true, allcolors=blue]{hyperref}
\usepackage{makecell}

\title{HARMONI at ELT: optical design of the spectrograph sub-system}

\author[a]{E.\ R.\ Muslimov}
\author[a]{M.\ Tecza}
\author[a]{E.\ Castillo-Dominguez}
\author[a]{J.\ Kariuki}
\author[a]{L.\ Boland}
\author[a]{M.\ E.\ Cisneros-Gonzalez}
\author[a]{K.\ McCall}
\author[a]{S.\ Paszynska}

\affil[a]{University of Oxford, Department of Physics, Denys Wilkinson Building, Oxford, OX1 3RH, UK}

\authorinfo{Further author information: (Send correspondence to E.R.M.)\\E.R.M.: E-mail: eduard.muslimov@physics.ox.ac.uk}

\begin{document}
\maketitle

\begin{abstract}
HARMONI is an adaptive optics assisted, near infrared integral field spectrograph for the European Extremely Large Telescope. It covers a spectral range from 750\,nm to 2450\,nm with resolving powers from 3000 to 7000 and spatial sampling of 25\,mas and 6\,mas. It can operate in two adaptive optics modes, SCAO and MCAO. The spectrograph represents the last optical sub-system of the HARMONI instrument optical train that disperses the light and forms a two-dimensional spectrogram on its scientific detectors. Changes in the top-level goals and specifications have led to a significant revision of the spectrograph optical design. The current baseline design consists of a two-mirror unobscured collimator, a set of six  volume-phase holographic grisms, and a two-mirror unobscured camera. The design is more compact than the previous version, it uses anamorphic freeform mirrors to correct aberrations, and is expected to reach a minimum transmission 57\% with improved accessibility of the detector unit and reduced scattered light.
\end{abstract}


\section{Introduction}
\label{sec:intro}
HARMONI is a key instrument for the coming 39-m European Extremely Large Telescope that will provide integral field spectroscopy over a wide spectral range. The initial set of requirements and instrument structure were developed in order to cover a long list of science cases. As a result,  HARMONI had an impressive number of operation modes and configurations and a complex hierarchy of systems and subsystems. In 2024-25 the instrument consortium has taken an effort of a deep design re-scope in order to reduce the instrument complexity and meet the existing constrains. A high priority was given to increase of the transmission and pointing stability, simplification of the instrument alignment and excluding single-point failure risks as those associated with cryogenic mechanisms. In addition, it was very desirable to reduce the mass and volume of the instrument.  \cite{MacIver2026}.

The scientific spectrograms are formed by the Integral Field Spectrograph (IFS) system. Its optical train includes the following sub-systems: IFS Pre-Optics (IPO), Integral Field Unit (IFU), IFS Spectrograph (ISP) and Infrared Science Detector Module (ISDM). For the IFS the design rescope implied a few first-order changes:
\begin{enumerate}
    \item Instead of four platescales $60mas\times30 mas$, $20mas\times20 mas$, $10mas\times10 mas$ and $4mas\times4 mas$ the instrument will operate with two platescales : $25mas\times25 mas$ and $6mas\times6 mas$.
    \item The visible and four high-resolution configurations are removed, so the system will need only two low-resolution (LR) and four medium-resolution (MR) dispersers. At the same time, the working range of the shortwave LR dispersers was extended down to 750 nm in order to match the updated interface with the telescope.
    \item The highest priority is given to elimination of moving parts and mechanisms.
    \item The top goal height reduction for the system is $\approx 500 mm$.
\end{enumerate}

Below we describe the design decisions made for the spectrograph sub-system (ISP) in context of these rescore priorities and present estimations for some key performance parameters. The rest of the paper is organized as follows:  Sec.~\ref{sec:design} presents the overall design concept and more detailed description of the mirror and disperser models; Sec~\ref{sec:perf} discusses the nominal performance including the spectral resolving power and transmission as well as compensation of the under-sampling effect; Sec~\ref{sec:conc} provides the conclusions and outlines the plan for future work.

\section{Optical Design}
\label{sec:design}
In attempt to account for all the re-design goals listed above we had to implement a deep re-design of the spectrograph optics. It still consists of three basic modules - collimator, disperser and camera, but each of the modules is notably different from its version in the previous design \cite{Thatte2020}. Moreover, the approach for design and performance assessment has been changed as well.  

\subsection{Design Approach}

Reducing the coarse platescale to $25 mas \times 25 mas$ implies that the maximum equivalent Numerical Aperture at the ISP input becomes $NA=0.0377$. Since the spaxel (spatial pixel) corresponds to $2\times1$ detector pixels, the geometrical pupil represents an ellipse with aspect ratio of $0.5$, where the short semi-axis corresponds to the spectral dimension, and the long semi-axis -- to the spatial one. 

Together with the intention to increase the spectrograph transmission and reduce its volume, this motivates us for use off-axis all-reflective designs for the collimator and camera unit. An unobscured geometry described via offset of field or pupil inevitably introduces aberrations such as coma and astigmatism. In order to correct these aberrations we have to add high-order aspheric terms to the mirror surface profiles. This goal could be reached with off-axis fraction of a coaxial design, where the mirror surfaces have axial symmetry, but the surface vertex does not coincide with the aperture centre\cite{Li2021}. Alternatively, each mirror can represent a freeform surface with tangential plane (\textit{YZ}) symmetry only, where the mathematical vertex is located at the centre of the part \cite{Bauer21}. We prefer the latter definition since this allows us to use the real and reachable vertices as optical and mechanical references. 

The spectrograph input represents a pseudo-slit, i.e. a set of $6.63mm \times0.13 mm$ slitlet images arranged in a checkerboard pattern across the $541.6 mm \times8mm $  exit focal plane of the IFU. The output focal plane should fit the science detector that represents an assembly of two HAWAII-2RG MCT sensors\cite{Hanaoka2020}, so the active zone is $125.52 mm\times 61.08 mm$ and the pixel size is $15 \mu m$. With these interfaces we can keep the optics size reasonably small and reduce the maximum angular field at the intermediate pupil plane coinciding with the disperser to$\approx 10^\circ$. It is known from the design practice that for this combination of field and aperture it is possible to use only two freeform surfaces to reach a good aberrations correction. For both of the imaging modules, collimator and camera, it is reasonable to use an approach similar to the classical Schmidt design, where the primary component is placed next to the pupil and introduces the aberration correction and the secondary component is approximately concentric to it and carries the most of the system's optical power. The classical solution cannot be applied directly because of the disperser position and boundary conditions defining the clearances in the system. In addition, the collimator receives a telecentric beam, although the output beam after the camera can deviate from the exact telecentricity.

For the disperser unit the goal of removing the beam folding mechanisms and the decrease of the pupil linear size give a strong motivation to switch to a grism geometry, where the incidence and diffraction angles for all the dispersers remain the same while the change for different wavebands is compensated with prisms. Otherwise, the dispersers are still based on volume-phase holographic (VPH) gratings that can have high diffraction efficiency in the transmitted $-1^{st}$ diffraction order.

\begin{figure}[ht]
\begin{center}
\includegraphics[width=\linewidth,height=0.48\textheight,keepaspectratio]{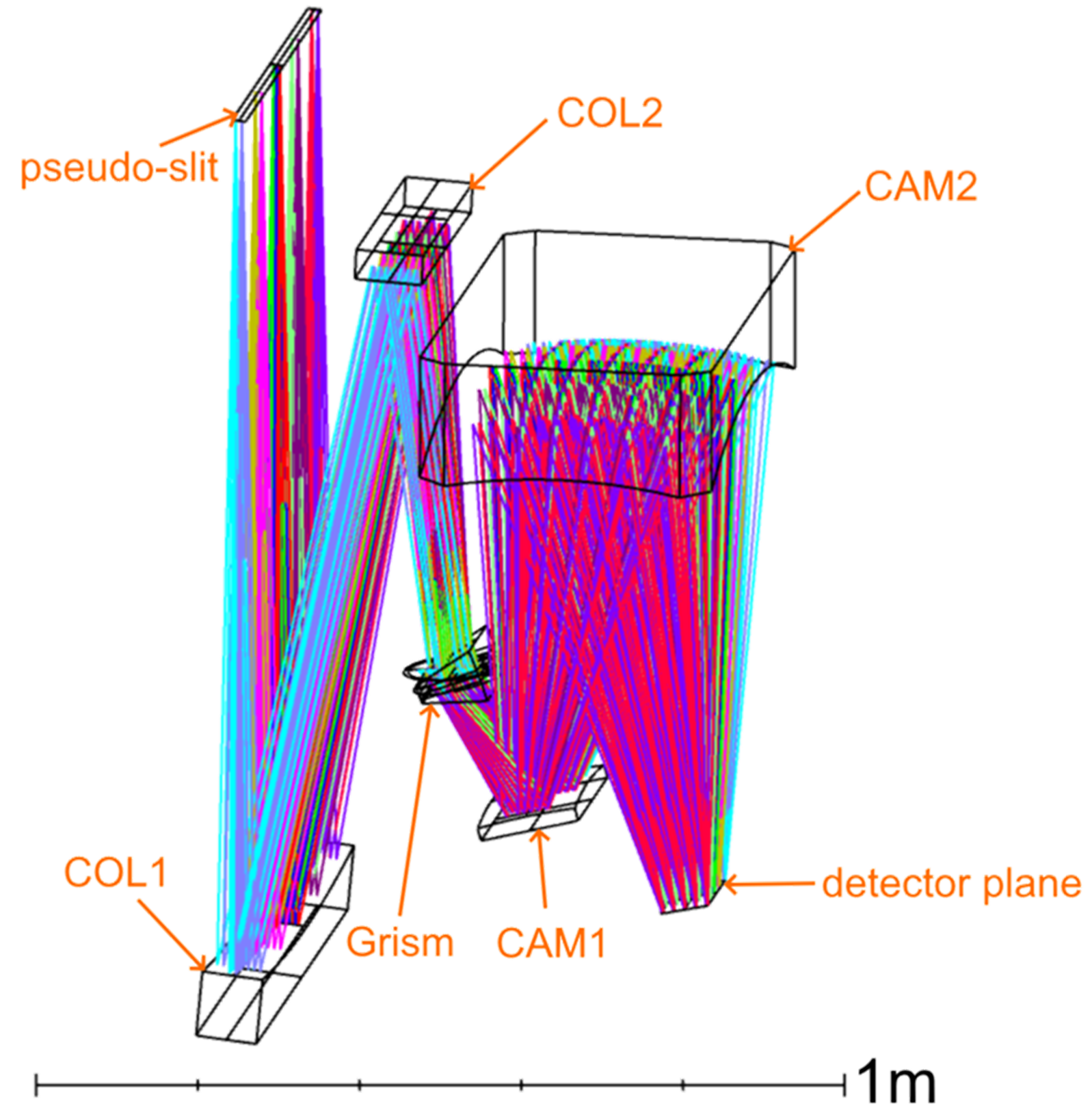}
\end{center}
\caption{\label{fig:layout_only} HARMONI spectrograph optical design layout.}
\end{figure}

The resultant optical design is shown in Fig.~\ref{fig:layout_only}. The overall dimensions are $1164 mm \times 646 mm \times 650 mm$. This figure includes realistic optical components with technological margins, but does not yet include the ISP mechanics.

\begin{figure}[ht]
\begin{center}
\includegraphics[width=\linewidth,height=0.38\textheight,keepaspectratio]{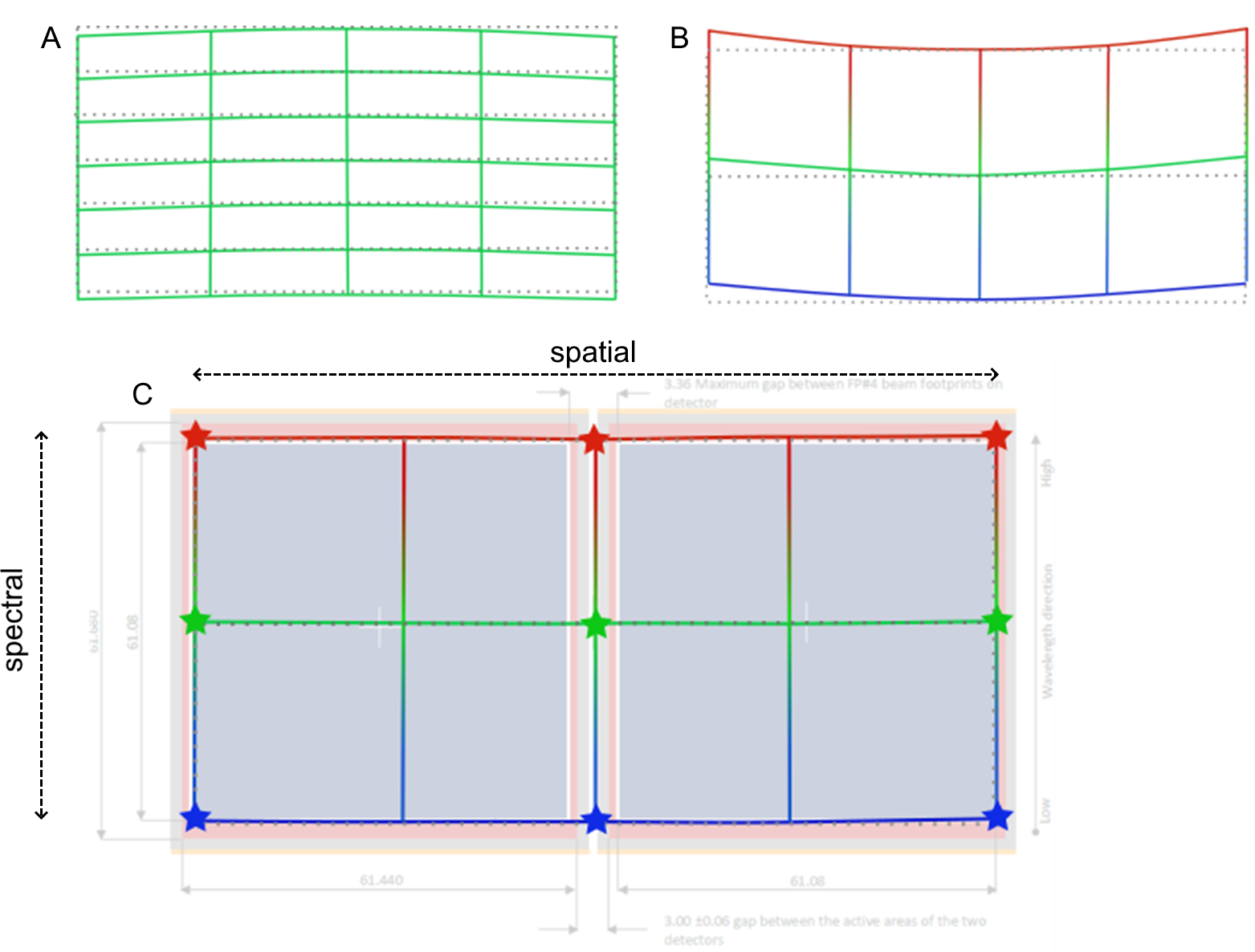}
\end{center}
\caption{\label{fig:distortion_only} Distortion correction in the spectrograph: A -- camera keystone distortion, B -- grism smile distortion, C -- compensated distortion for MR1 configuration shown on the scale of science detector.}
\end{figure}

A final remark here should be made regarding the distortion compensation. The collimator distortion is relatively low ($\approx0.4\%$), while the camera distortion is significant $\approx2.5\%$ and has an inherent "keystone" pattern -- see Fig.~\ref{fig:distortion_only},A. On the other hand, the grism introduces a distortion in the spectral direction that has "smile" pattern, so compensates the camera's distortion (Fig.~\ref{fig:distortion_only},B). This distortion varies with the dispersion and equals to $6.1\%$ for LR dispersers and $2.5\%$ for the MR ones. Therefore, the mutual compensation of these effects can be used to increase the usable height of the detector and improve the spectral resolution (Fig.~\ref{fig:distortion_only},C). Moreover, since the linear magnification in the spectral direction must be controlled to a high precision, it becomes necessary to cross-verify the distortion compensation. In our case, this means applying two optimization modes (i.e. set of targets, variables and boundary conditions) in a loop to the same model that includes both the dispersers and the camera.       

\subsection{Freeforms Design}

As we stated above, the aberrations correction requires using freeform surfaces and we prefer to keep an accessible surface vertex, so each mirror represents a freeform tilted with respect to the X axis passing through its vertex. 

The entire design is inherently anamorphic, since the pupil is elliptical, the output focal plane format is rectangular and the linear magnification must be different in the tangential and sagittal planes. So, it becomes quite natural to use an anamorphic equation as the base profile for the freeform mirrors. In our case, we use the biconic surface equation, defined by two vertex radii of curvature $R_x$ and $R_y$ and the corresponding conic constants $k_x$ and $k_y$.   

\begin{figure}[ht]
\begin{center}
\includegraphics[width=\linewidth,height=0.38\textheight,keepaspectratio]{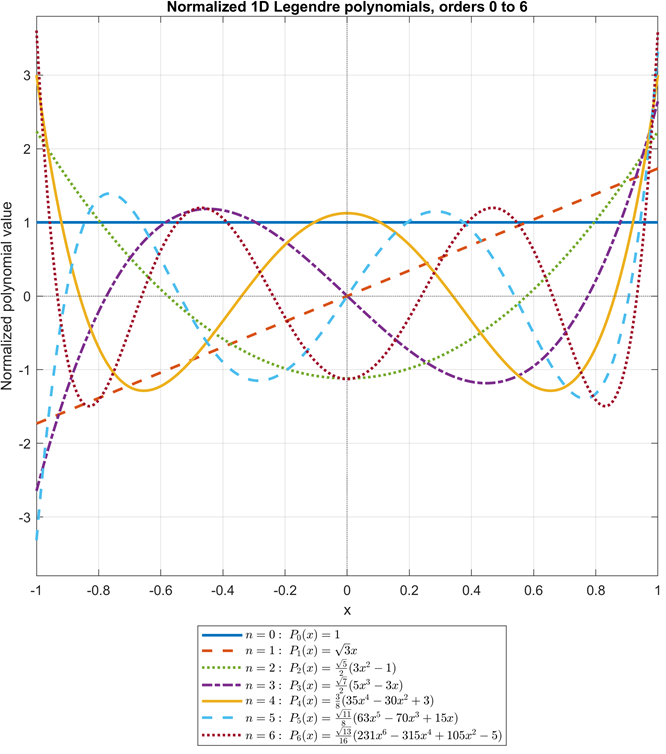}
\end{center}
\caption{\label{fig:legendre_basis} One-dimensional Legendre basis functions used in the freeform parameterization.}
\end{figure}

On top of this, for correction of higher-order aberrations we introduce XY polynomials up to the $7^{th}$ order. It was shown before \cite{Nikolic2016, Ye2014} that orthonormal polynomials show better convergence and it may be beneficial to use the polynomials defined on rectangle when the clear aperture is nearly rectangular. For this specific task, we use Legendre polynomials. The first 7 polynomials for one variable are shown in Fig.~\ref{fig:legendre_basis}. The 2D version of them represents just a linear combination of polynomials for X and Y. 

Thus, each of the four freeform surfaces is described by the following equation:
\begin{equation}
\label{eq:freeform_surface}
S(X,Y)=
\frac{\dfrac{X^2}{R_x}+\dfrac{Y^2}{R_y}}
{1+\sqrt{1-\dfrac{(1+k_x)X^2}{R_x^2}-\dfrac{(1+k_y)Y^2}{R_y^2}}}
+ \sum_{j=1}^{7}\sum_{k=1}^{7} A_{jk}\,
P_j\!\left(\frac{X}{X_{\max}}\right)
P_k\!\left(\frac{Y}{Y_{\max}}\right),
\end{equation}

where the Legendre arguments are the coordinates on the surface normalized to the maximum $x=X/X_{\max}$ and $y=Y/Y_{\max}$. The first term in \ref{eq:freeform_surface} is the biconic sag, and the second term adds the two-dimensional Legendre correction with coefficients $A_{jk}$.  

\begin{figure}[ht]
\begin{center}
\includegraphics[width=\linewidth,height=0.45\textheight,keepaspectratio]{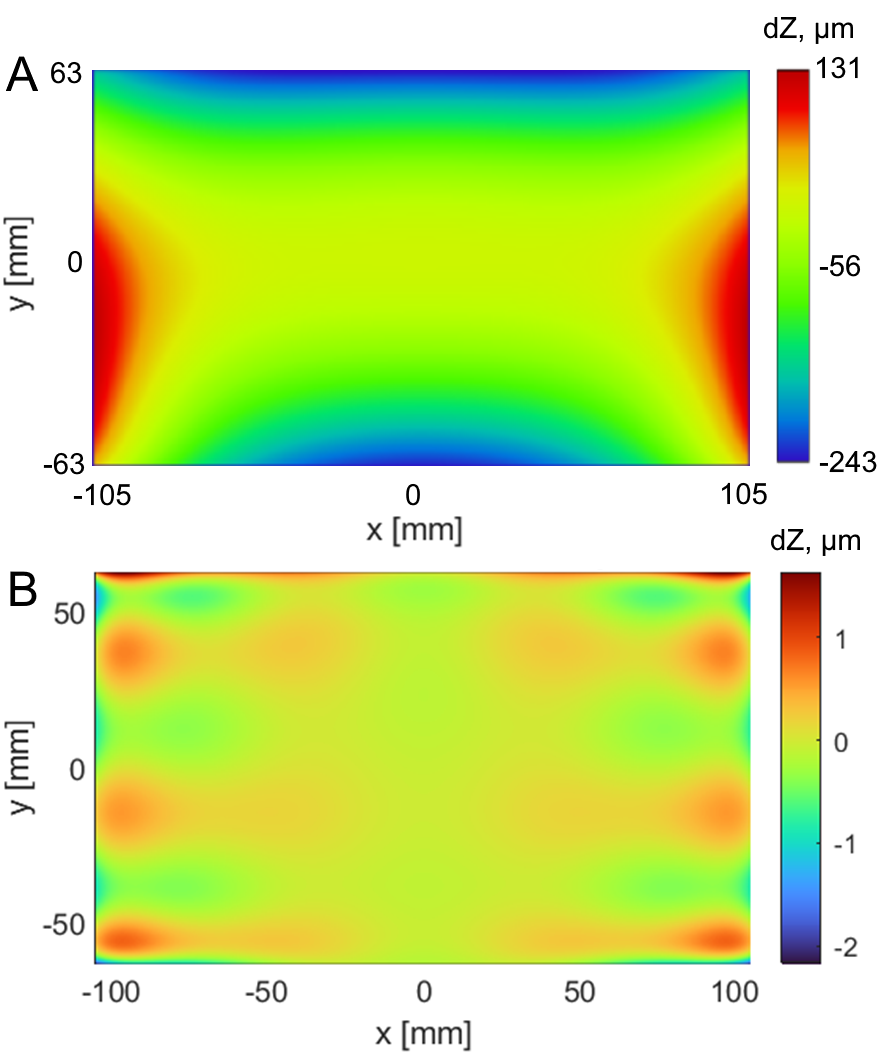}
\end{center}
\caption{\label{fig:freeform_impl} Examples of the freeform surface design: A -- aspheric departure in microns for the camera M1 mirror, B -- residual difference in microns for this surface defined via Zernike and Legendre polynomials.}
\end{figure}

The obvious downside of this approach is the absence of standard tools to model such a surface in an optical design software. However, implementation of a user defined surface is relatively straightforward. Fig.~\ref{fig:freeform_impl},A demonstrates the sag deviation from the best fit sphere calculated for CAM-1 mirror. The Legendre polynomials help to increase the optimization convergence, but they can be added at later stages. They can be converted to more commonly used Zernike polynomials \cite{Ye2014} using a numerical fitting method, e.g. simplex method \cite{Nelder65}. The residuals between the two sets of orthonormal polynomials for the same  CAM-1 mirror are shown in Fig.~\ref{fig:freeform_impl},B. 

\subsection{Grisms Design}

Each of the six dispersers represents a grism, so it consists of a VPH grating protected with a cover glass and capped by two identical prisms. Using two prisms helps to moderate the prism angle and maintain clamping forces symmetry in a mounted grism. All of the elements are made of fused silica.   

However, the grisms design cannot be called symmetrical in the exact meaning of this term. There are two reasons to deviate from a perfect symmetry. 

\begin{figure}[ht]
\begin{center}
\includegraphics[width=\linewidth,height=0.48\textheight,keepaspectratio]{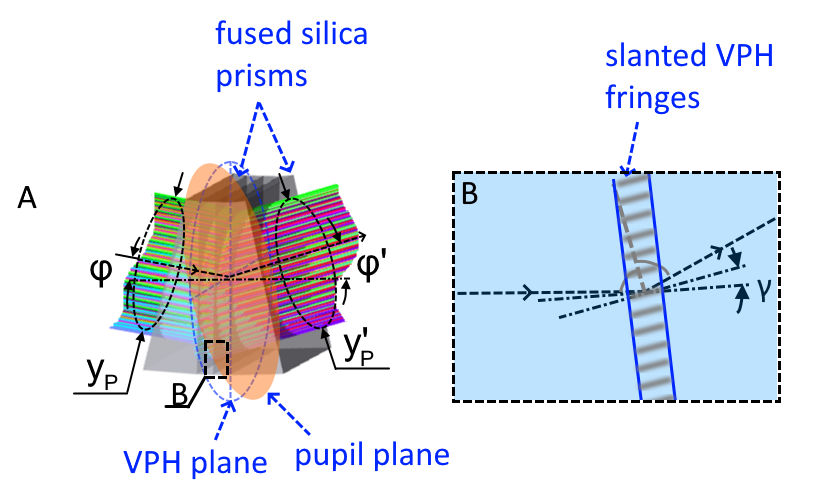}
\end{center}
\caption{\label{fig:grisms} Grism disperser design on the MR1 example: A -- basic geometry with anamorphic beam conversion, B -- slanted VPH fringes geometry.}
\end{figure}

First, the IFS is expected to suffer from a spectral under-sampling caused by a spatial filtering of the diffracted pupil --see in detail in \cite{Todd2026}. In order to compensate this undesirable effect we should introduce an anamorphic magnification for each monochromatic beam that would stretch each slitlet image by a factor proportional to the wavelength. The most straightforward to introduce such an anamorphism is to deviate from a perfectly symmetrical diffraction geometry, so the incidence $\phi$ and diffraction $\phi'$ angles becomes unequal and the magnification is 
\begin{equation}
\label{eq:anamr}
\beta_y\sim  \frac{y_P}{y'_P}\sim  \frac{cos(\phi)}{cos(\phi')}.
\end{equation}
See Fig.~\ref{fig:grisms},A for the variables notation.

 Another reason for having asymmetric diffraction geometry at the VPH is removing the Littrow ghost. If a grating has perfectly symmetrical diffraction geometry, the optical paths in two propagation directions coincide, and the light reflected from the sensor can create a strong recombination ghost \cite{Burgh2007}. So, the VPH fringes should be slanted to satisfy the Bragg condition for such an asymmetrical diffraction geometry -- see Fig.~\ref{fig:grisms}, B.

 Otherwise, the targets driving the grisms design are quite obvious. Each of them has a working spectral sub-band defined by the atmospheric transmission and the science goals. For each grism the spectral image must match the detector active area size.  All the geometrical parameters for grisms are summarized in Table~\ref{tab:grisms}.

\begin{table}[ht]
\caption{\label{tab:grisms} Nominal design parameters of the ISP grisms}
\begin{center}
\scriptsize
\resizebox{\linewidth}{!}{%
\begin{tabular}{|l|c|c|c|c|c|c|}
\hline
Parameter & LR1 & LR2 & MR1 & MR2 & MR3 & MR4 \\
\hline
Min wavelength (nm) & 750 & 1450 & 830 & 1046 & 1450 & 1970 \\
Central wavelength (nm) & 1059.5 & 1950 & 940 & 1185 & 1625 & 2210 \\
Max wavelength (nm) & 1369 & 2450 & 1050 & 1324 & 1800 & 2450 \\
Prism angle (deg) & -15.305 & -13.254 & 13.312 & 13.256 & 14.710 & 14.131 \\
Inc.\ angle in air (deg) & 30.246 & 24.972 & 3.588 & 3.578 & 13.720 & 13.021 \\
Exit.\ angle in air (deg) & 32.490 & 32.527 & 32.763 & 32.759 & 32.713 & 32.702 \\
Lines density (1/mm) & 279.093 & 172.869 & 818.095 & 647.002 & 476.117 & 344.266 \\
Slanting angle (deg) & 0.822 & 2.924 & 4.580 & 4.620 & 10.581 & 10.555 \\
Full aperture X (mm) & 127 & 127 & 127 & 127 & 127 & 127 \\
Full aperture Y (mm) & 80 & 78 & 73.4 & 73.4 & 76 & 76 \\
\hline
\end{tabular}}
\end{center}
\end{table}

\section{Performance Estimation}
\label{sec:perf}
This section provides a quick overview of the key performance metrics of the spectrograph. Note that these estimations are preliminary since the design parameters are not yet corrected for the technological limitations and necessary margins for finite manufacturing and alignment errors are not taken into account. We introduce safety factors for these effects, where possible. 

\subsection{Spectral Resolving Power}

The spectral resolving power is defined as follows:
\begin{equation}
\label{eq:resol}
R=\frac{\lambda}{\Delta \lambda}=\frac{\lambda}{\beta_y f'_{cam}/f'_{col} \partial y'/\partial \lambda FWHM 
}.
\end{equation}
Here $\lambda$ is the wavelength, $\Delta\lambda$ -- spectral resolution, $\beta_y$ -- grism anamorphic magnification,   $f'_{cam}$ -- camera focal length, $f'_{col}$ -- collimator focal length, $\partial y'/\partial \lambda $ -- reciprocal linear dispersion and $FWHM$ is the full width at half maximum of the instrument function that, in turn, represents a convolution of the geometrical image intensity (a rectangular function) with the line-spread function (LSF) of the ISP optics.

\begin{figure}[ht]
\begin{center}
\includegraphics[width=\linewidth,height=0.43\textheight,keepaspectratio]{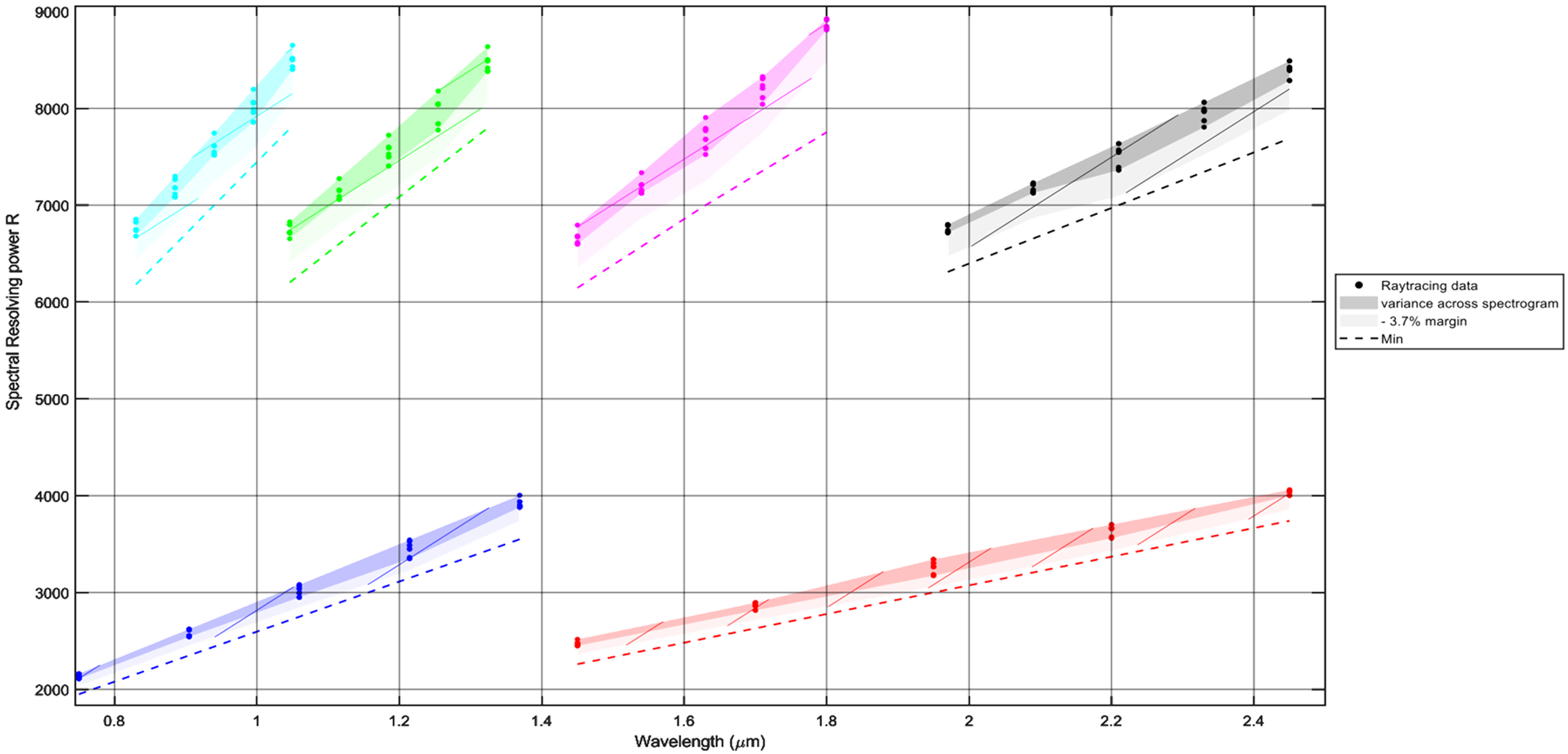}
\end{center}
\caption{\label{fig:resolving_power} Spectral resolving power summary for the current design set.}
\end{figure}

 The resolving power values calculated for the nominal state of ISP, i.e. before adding any manufacturing and assembly errors, are summarized in Fig.~\ref{fig:resolving_power}. The dashed lines correspond to the sub-system requirements, while the individual points represent the $R$ calculated for different spatial field positions. Shaded zones indicate the spatial variance and the hatched zones show the expected degradation in the "as-built" state ($\approx4\%$). As one can see, the safety margin for the nominal design counts to $\approx5-10\%$, so the actual as-built spectrograph should meet the specifications.

\subsection{Transmission}

Transmission was one of the key drivers behind the instrument design rescope. With the current design we expect a notable gain in transmission for the following reasons:
\begin{enumerate}
    \item The collimator and camera optics are all-reflective. This excludes the K-band absorbance and bulk scattering issues that limited the performance in the previous design version.
    \item The collimator and camera optics in total have only four surfaces. This means that we can expect a high and  uniform transmission with protected silver coating ($\rho\approx98.5\%$).
    \item We can also expect a low scattering level, since the mirrors can be polished to a low roughness $Ra= 2..5nm$ and the particle and molecular contamination can be low. Two out of four mirrors, including the largest one, are facing downwards that should help reducing the cumulative particle contamination.
    \item All of the grisms can be made with fused silica substrates and prisms that have high transmission over the entire spectral band.
    \item In the current design there are no refractive surfaces working for the entire spectral range. This means that the anti-reflective (AR) coatings can be optimized for each sub-band and reach a better performance with the residual reflectivity of $\rho\approx0.5\%$.
    \item The recent measurements of VPH gratings scattering \cite{Jeanneau24} show that the scattering losses can be very low.
\end{enumerate}

\begin{figure}[ht]
\begin{center}
\includegraphics[width=\linewidth,height=0.43\textheight,keepaspectratio]{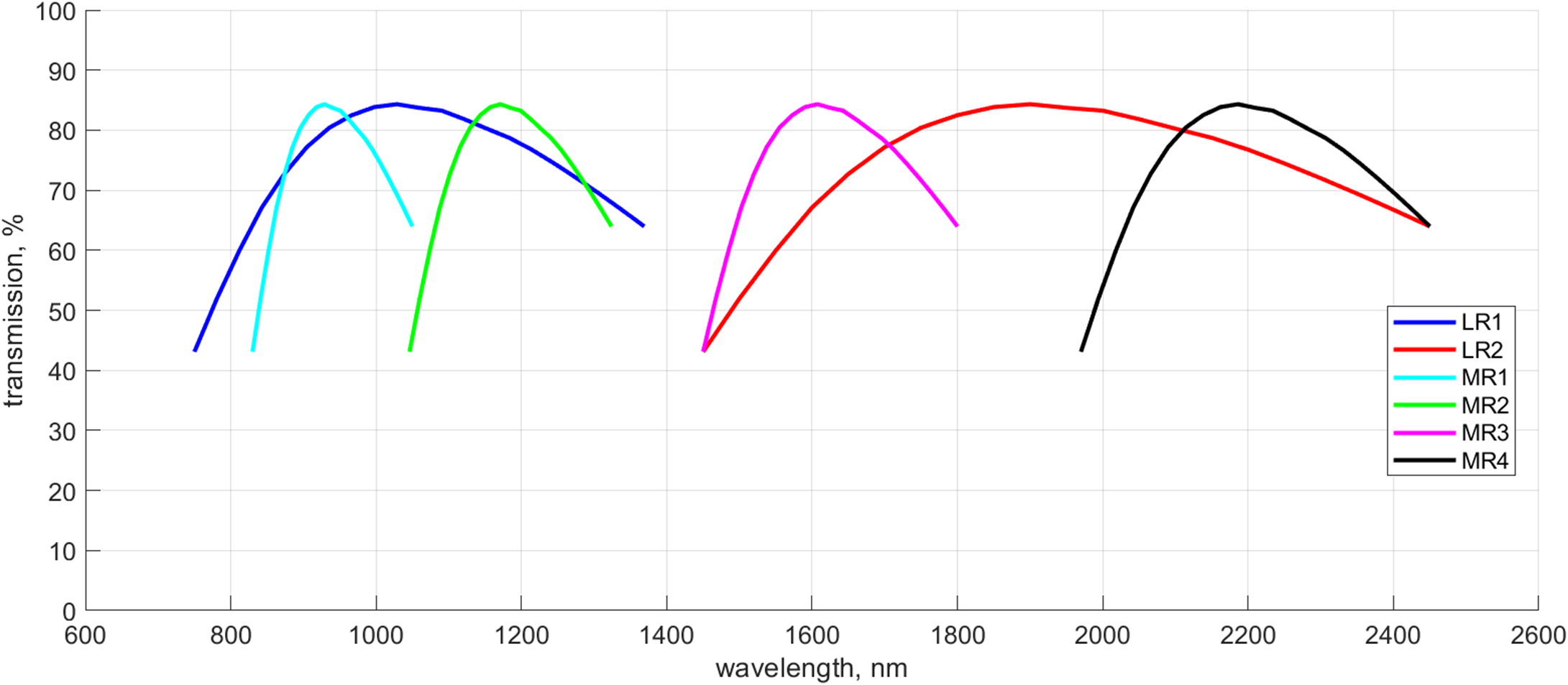}
\end{center}
\caption{\label{fig:transmission} Estimated transmission for all the spectrograph configurations.}
\end{figure}

With all of the above-mentioned points, the VPH grating diffraction efficiency becomes the driving factor behind the transmission budget. For a preliminary estimation we use the Rigorous Coupled Waves Analysis method \cite{Moharam81} to calculate and optimize the diffraction efficiency. A margin of $7\%$ is allocated for the losses in the VPH grating. It is presumed that the VPH's are recorded in dichromated gelatin (DCG)\cite{Naik1990}, although it can be made with a different material. The estimated transmission across the six working spectral sub-bands are shown in Fig.~\ref{fig:transmission}. The spectrograph requirements exclude the very edges of each sub-band and set target values for the central $80\%$. The minimum values for these reference ranges varies from $57.0\%$ to $74.2\%$, while the corresponding average values are $74.7-81.1\%$.

\subsection{Anamorphic Magnification}

In the context of under-sampling issue revealed in \cite{Todd2026}, the anamorphic magnification compensating for the diffracted pupil cropping effect becomes one of the key performance metrics for the spectrograph sub-system. It creates an obvious contradiction with the resolving power requirement since increase of the$\beta_y$ magnification in Eq.~\ref{eq:anamr} leads to decrease of the resolving power according to Eq.~\ref{eq:resol}. Further, the under-sampling effect scales with the wavelength, so the target anamorphic magnification changes across each sub-band and from one sub-band to another. One more factor affecting the anamorphic factor is the grating dispersion. It is easy to see by differentiating the well known grating equation that the anamorphic magnification grows with the wavelength, since
\begin{equation}
\label{eq:disp}
\phi'=asin(k \lambda N -sin(\phi))
\end{equation}

where $N$ is the grating lines density and $k$ is the diffraction order. 
At the same time, the residual distortion shown in Fig.~\ref{fig:distortion_only} will also affect the magnification across the spectral dimension.
Finally, all of these factors will experience some variance in the as-built instrument due to the field aberrations and manufacturing and alignment errors.

\begin{figure}[ht]
\begin{center}
\includegraphics[width=\linewidth,height=0.48\textheight,keepaspectratio]{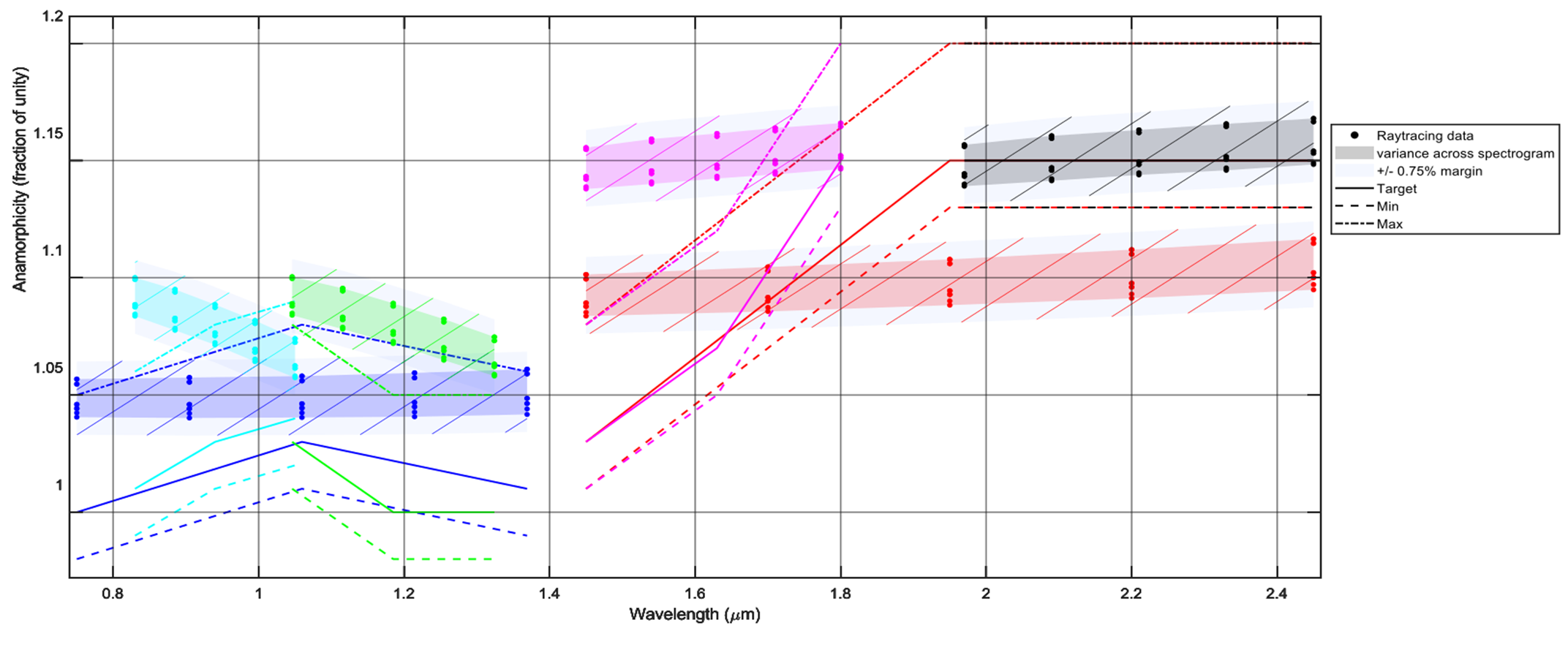}
\end{center}
\caption{\label{fig:anamorphism} Anamorphic magnification summary used in the performance trade study.}
\end{figure}

The current estimates for the anamorphic magnification are shown in Fig.~\ref{fig:anamorphism}. Each dot represent the anamorphic magnification in the nominal design for a single field position, so the coloured bands show the spatial variance. The hatching indicate the expected variance in the as-built state, where the errors add about $1.5\%$ on top of the change in the nominal design. The solid curves correspond to the target values, while the dashed ones show the tolerances assigned for the anamorphic magnification.

The plot indicates that it is practically impossible to follow the target curves exactly because of all the boundary conditions listed above. In most of the cases this leads  to a slight over-stretching of the monochromatic image. This effect is acceptable as long as the resolving power meets the  requirements (see Fig.~\ref{fig:resolving_power}). However, at longer wavelengths $>2 \mu m$ it becomes difficult to match the required anamorphic factor since it would lead either to losses in the resolving power, or to technologically risky solutions as large VPH fringes slanting angles. Therefore the anamorphic magnification values was set to the highest reachable value for this part of the working band.   

\subsection{Image quality metrics comparison}

In a perfect case the image quality metric should meet the following criteria:
\begin{enumerate}
    \item It should be traceable through the entire development process -- design, modeling, parts and modules manufacturing, alignment and test of the entire sub-system;
    \item It should be measurable directly in an experiment and independent of the data post-processing techniques;
    \item It should  cover all the wavelengths and for the entire field of view;
    \item It should represent the eventual on-sky performance in order to avoid over- or under-estimating the actual performance through indirect measurements.
\end{enumerate}

One approach commonly used in instrumentation design implies using wavefront error (WFE) as the main criterion at all levels from individual surfaces to the entire optical system. This approach allows for a simple root square sum break-down of the image quality budget. It has some obvious limitations such as the necessity to use a certain interferometer laser wavelength or difficulty of scanning through a large field of view. For a cryogenic instrument as HARMONI measurement of the WFE also creates some difficulties related to coupling the interferometer beam into the optical system mounted inside of a cryo-chamber. Also, for the specific case of HARMONI spectrograph, WFE can underrepresent its performance since it is naturally insensitive to any spectral/spatial asymmetry. On top of this, it appears that  changes in the WFE do not represent the changes in resolving power, since small aberrations affect the LSF wings, but do not change the FWHM. In order to quantify the latter effect we calculated the root mean square (RMS) WFE and the spectral FWHM for a number of wavelengths and field positions in the nominal optical design that cover all the configurations and the whole spectrogram.

Then we calculated the Pearson linear correlation coefficient ~\citenum{Benesty09} over this dataset. The correlation coefficient for any pair of variables $x$ and $y$ equals 
\begin{equation}
\label{eq:pearson}
r_{xy}=
\frac{\sum_{i=1}^{n}(x_i-\bar{x})(y_i-\bar{y})}
{\sqrt{\sum_{i=1}^{n}(x_i-\bar{x})^2}\sqrt{\sum_{i=1}^{n}(y_i-\bar{y})^2}},
\end{equation}
where $r_{xy}$ ranges from $-1$ to $1$ and measures the strength and direction of the linear relationship between two sampled variables.

\begin{figure}[ht]
\begin{center}
\includegraphics[width=\linewidth,height=0.48\textheight,keepaspectratio]{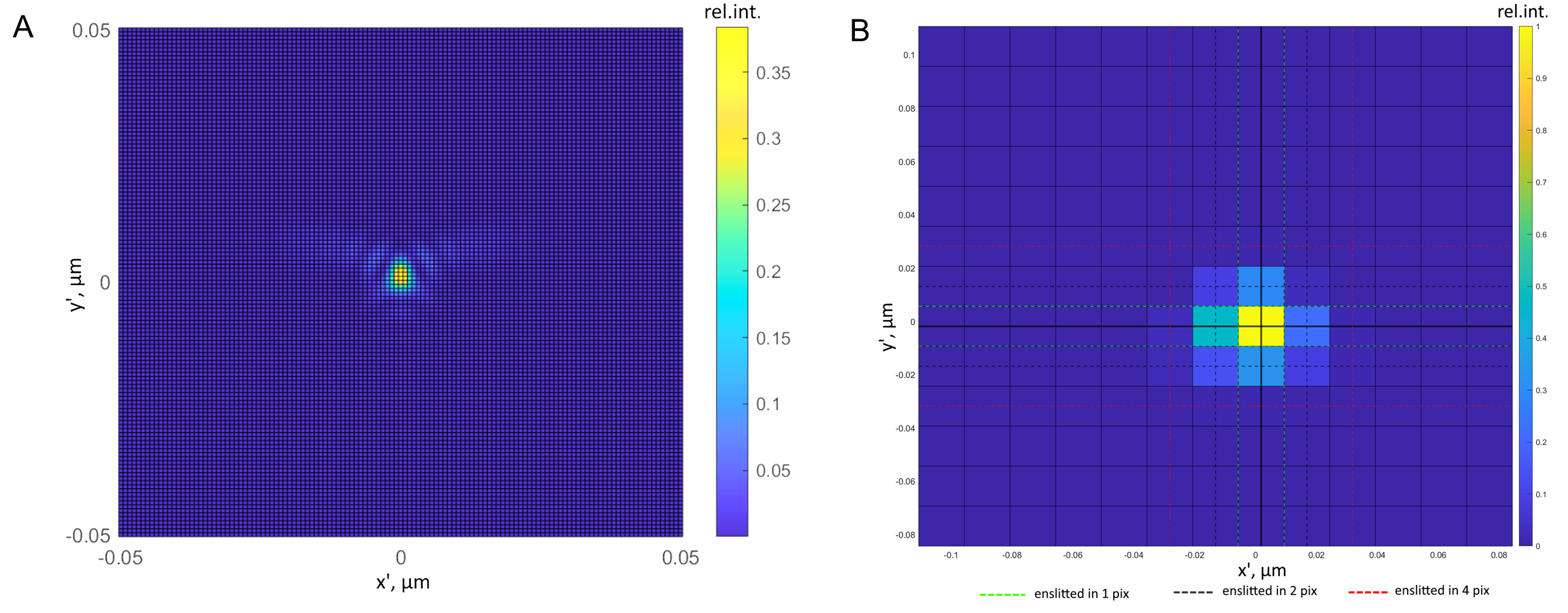}
\end{center}
\caption{\label{fig:psf_fiber_only} PSF and fiber-image illustration used in the metrics comparison.}
\end{figure}

Then we compared this coefficient with a few other image quality metric. First, we calculate the geometrical spot RMS radii. This is a metric commonly used for design and optimization, but unfortunately it cannot be measured directly in an experiment, especially for systems approaching the diffraction limit. Then, we calculate the aberrated point spread function (PSF) of the spectrograph (see Fig.~\ref{fig:psf_fiber_only}, A). Similarly, for a high resolution application there is no practical way to neglect the finite size of the object, so it is impossible to measure PSF directly. So we propose to detect images of a multimode fiber end. Its image will represent a convolution of uniformly bright circle of $105 \mu m$ in diameter and the PSF. In a practical test one can set up a few fibers to cover the entire spatial field and use a relay with intermediate aperture stop to represent the input beam aperture. Each wavelength across the working range could be probed separately by use of a monochromator feeding light into the fiber. The resultant image can be detected using the spectrograph's science sensor, so it will be sampled with the $15 \mu m$  pixels. An example of such an image simulation is shown in Fig.~\ref{fig:psf_fiber_only},B. Then in order to quantify the performance separately for the spatial and spectral directions we calculate the fraction of total energy enslitted in the rows/ columns of 1-4 pixel in width centered with respect to the maximum (see the dashed lines in Fig.~\ref{fig:psf_fiber_only},B).

Table~\ref{tab:correlations} summarizes the linear correlation coefficients for all of the image quality metrics discussed above. The extreme cases for the FWHM are highlighted. The values demonstrate that it is very hard to predict the spectral resolving power performance by the RMS WFE measurements. Typically this means that large change in the WFE leads to only a mild increase of the FWHM.  So, there is a risk over-constraining the WFE values that, in turn, can lead to challenging tolerances and long and expensive manufacturing and alignment process. On contrary, the  energy enslitted in the row containing the brightest pixel shows a very strong negative linear correlation with the FWHM value, which is intuitively clear. Notably, the 4-pixel Y enslitted energy also shows a strong correlation, but the value calculated for the 2 pixel width is much less informative.

\begin{figure}[ht]
\begin{center}
\includegraphics[width=\linewidth,height=0.4\textheight,keepaspectratio]{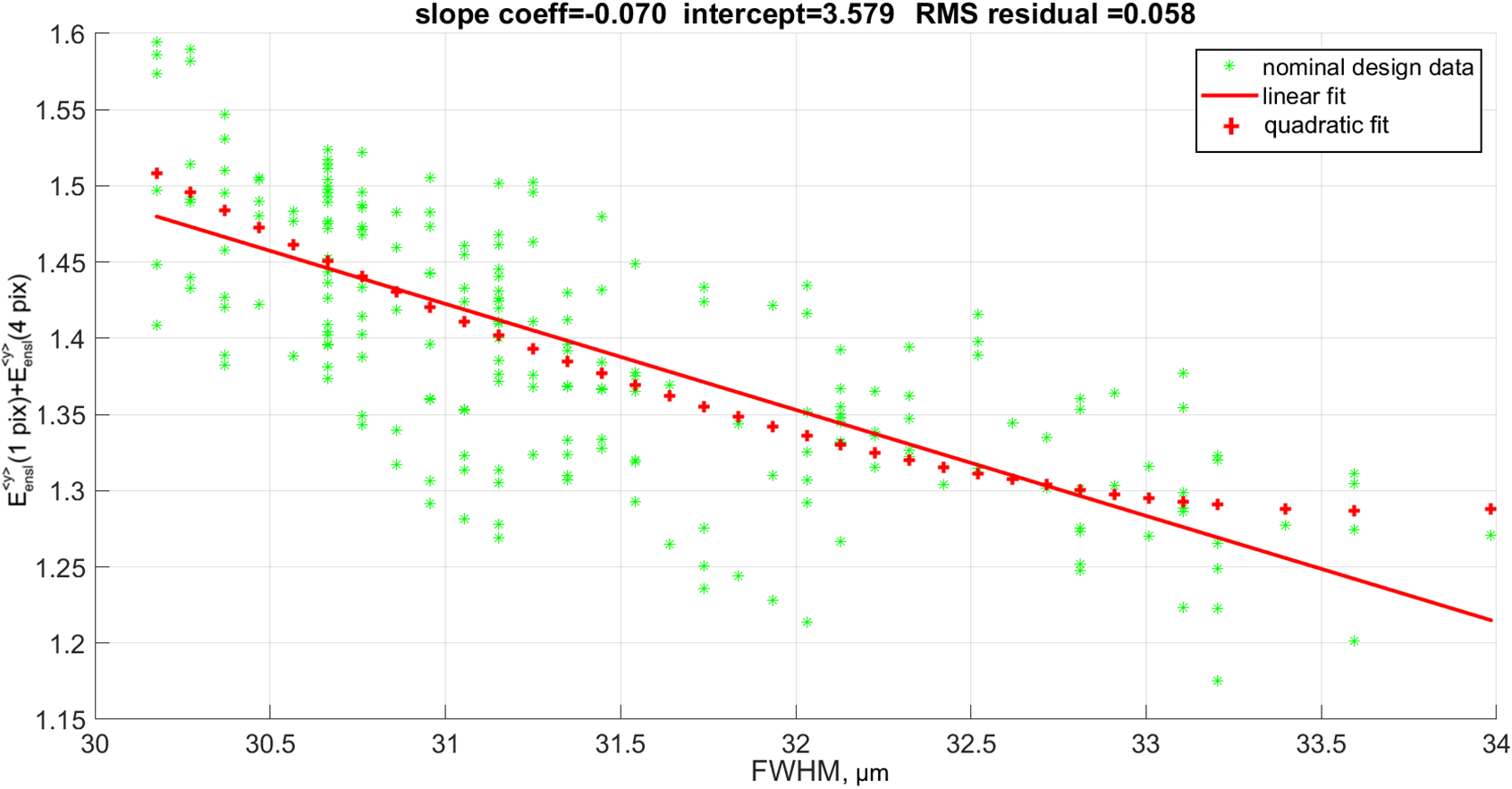}
\end{center}
\caption{\label{fig:enslitted_metric_only} Covariance of the proposed enslitted energy metric and the FWHM overlaid with the linear and quadratic fit curves. }
\end{figure}

Apparently, it is possible to increase the linear correlation even more by using the simplest synthetic metric equal to $E_{ensl}^{<y>} (4 pix)+E_{ensl}^{<y>} (1 pix)$. In this case the correlation coefficient grows to $0.74$. Fig.~\ref{fig:enslitted_metric_only} below demonstrates the distribution of this metric calculated across the spectral range and spatial field versus the  FWHM values. The correlation is apparent and the red line visualizes the linear fit. For comparison the dotted curve shows the quadratic fit and it does not lead to a significant decrease in the residual fitting error. So, it is the linear dependence between the two values can be used for practical purposes. It is important to note that the LSF in these calculations was over-sampled, so changes in the FWHM much smaller than a pixel size become traceable. At the same time the enslitted energy metric does not require any complex computations like LSF curve fitting with limited sampling. 
\begin{table}[ht]
\caption{\label{tab:correlations} Correlation matrix for the spectrograph image quality metrics.}
\begin{center}

\renewcommand{\arraystretch}{1.5}
\resizebox{\linewidth}{!}{%
\begin{tabular}{|l|r|r|r|r|r|r|r|r|r|}
\hline
Metric & RMS WFE, nm & Spot RMS, $\mu m$ & FWHM, $\mu m$ & $E_{ensl}^{\langle y\rangle}$ (1 pix) & $E_{ensl}^{\langle y\rangle}$ (2 pix) & $E_{ensl}^{\langle y\rangle}$ (4 pix) & $E_{ensl}^{\langle x\rangle}$ (1 pix) & $E_{ensl}^{\langle x\rangle}$ (2 pix) & $E_{ensl}^{\langle x\rangle}$ (4 pix) \\
\hline
RMS WFE, nm & 1.000 & 0.653 & 0.080 & 0.020 & -0.132 & -0.045 & 0.198 & -0.054 & 0.051 \\
\hline
Spot RMS, $\mu m$ & 0.653 & 1.000 & 0.107 & -0.023 & -0.115 & -0.084 & 0.322 & -0.078 & 0.188 \\
\hline
FWHM, $\mu m$ & \textbf{0.080} & 0.107 & 1.000 & \textbf{-0.701} & -0.103 & -0.657 & -0.125 & 0.027 & -0.143 \\
\hline
$E_{ensl}^{<y>}$ (1 pix) & 0.020 & -0.023 & -0.701 & 1.000 & 0.160 & 0.693 & 0.427 & -0.178 & 0.278 \\
\hline
$E_{ensl}^{<y>}$ (2 pix) & -0.132 & -0.115 & -0.103 & 0.160 & 1.000 & 0.451 & -0.038 & -0.127 & 0.340 \\
\hline
$E_{ensl}^{<y>}$ (4 pix) & -0.045 & -0.084 & -0.657 & 0.693 & 0.451 & 1.000 & 0.189 & -0.237 & 0.615 \\
\hline
$E_{ensl}^{<x>}$ (1 pix) & 0.198 & 0.322 & -0.125 & 0.427 & -0.038 & 0.189 & 1.000 & -0.135 & 0.412 \\
\hline
$E_{ensl}^{<x>}$ (2 pix) & -0.054 & -0.078 & 0.027 & -0.178 & -0.127 & -0.237 & -0.135 & 1.000 & -0.354 \\
\hline
$E_{ensl}^{<x>}$ (4 pix) & 0.051 & 0.188 & -0.143 & 0.278 & 0.340 & 0.615 & 0.412 & -0.354 & 1.000 \\
\hline
\end{tabular}}
\end{center}
\end{table}
Thus, we believe that the approach based on the enslitted energy calculation is very promising for the spectrograph sub-system alignment and test. Besides the advantage of a good correlation with FWHM it can be easily applied for the spatial direction and also can be scaled to characterize the PSF/LSF wings, which are important for the high contrast mode.

\section{Conclusion}
\label{sec:conc}
In this paper we have presented the current state of the HARMONI spectrograph sub-system optical design. It allows us to meet the most of the goals of the instrument rescope study and to solve some issues found at the later stages of the development:
\begin{itemize}
    \item The design has a higher throughput;
    \item It meets the spectral resolving power requirements with a reasonably wide margin for manufacturing and alignment errors;
    \item It excludes the beam-folding mechanisms;
    \item This design has reduced the number of straylight sources;
    \item The current design is notably shorter than the previous version.   
\end{itemize}

There are also some other advantages as an easier access to the science detector module and potentially shorter production chain. 

Also, we have studied the imaging properties of this design and proposed a simple image quality metric that uses an enslitted energy fraction calculated for a multimode fiber end monochromatic image. 

Further steps in the optical system development should include:
\begin{itemize}
    \item Sensitivity analysis;
    \item Pre-assignment of tolerances and their correction to the technologically reachable values;
    \item Statistical analysis of the expected as-built performance based on the Monte-Carlo simulations including the alignment simulation;
    \item Ghosts analysis;
    \item Mechanical structure re-design;
    \item Estimation of the thermo-mechanical effects and their impact on the optical performance; 
    \item Analysis of the adjacent sub-systems impact, e.g. precision of the focusing mechanism, sensor flatness and chief ray deviations in the IFU output beam.
    
\end{itemize}

These activities should develop in parallel with prototyping of the key parts of the sub-system and interaction with the industry.

\section*{ACKNOWLEDGMENTS}       
This work is supported by UKRI-STFC grants  $\# ST/X002322/1$ and $ST/S001409/1$.

\bibliography{report}
\bibliographystyle{spiebib}

\end{document}